\documentclass[acmtog, nonacm]{acmart}
\usepackage{algorithmic} 
\usepackage{mathtools}
\usepackage{graphicx} 
\usepackage{textcomp} 
\usepackage{xcolor} 
\usepackage{hyperref} 

\usepackage{pgfplots} 
\usepackage{tikz}

\usepackage{balance} 

\usepackage{acronym}

\AtBeginDocument{%
  \fancypagestyle{plain}{%
    \fancyhf{} %
    \fancyfoot[L]{ACM SIGENERGY Energy Informatics Review}%
    \fancyfoot[R]{Volume 3 Issue 4, December 2022}}
  \providecommand\BibTeX{{%
    \normalfont B\kern-0.5em{\scshape i\kern-0.25em b}\kern-0.8em\TeX}}}
    
\let\oldmaketitle\maketitle
\renewcommand{\maketitle}{%
  \oldmaketitle%
  \thispagestyle{plain}%
  \pagestyle{plain}}

\newenvironment{datamaterial}%
{ \vspace{-0.15cm}%
    \small\noindent{\bfseries Availability of Data and Material:}\par%
    \noindent\ignorespaces}%
{ \par\noindent%
\ignorespacesafterend }%

\definecolor{colorred}{rgb}{1.0,0.0,0.0}
\definecolor{firstcolor}{HTML}{000000}
\definecolor{secondcolor}{HTML}{525252}
\definecolor{thirdcolor}{HTML}{737373}
\definecolor{fourthcolor}{HTML}{969696}
\definecolor{fifthcolor}{HTML}{bdbdbd}
\definecolor{sixthcolor}{HTML}{d9d9d9}
\definecolor{seventhcolor}{HTML}{f0f0f0}
\definecolor{eightcolor}{HTML}{ffffff}

\begin{document}

\title{Coupling OMNeT++ and mosaik for integrated Co-Simulation of ICT-reliant Smart Grids}
\author{Frauke Oest}
\authornote{F.O., E.F., and M.R. contributed equally to this research.}
\email{frauke.oest@offis.de}
\author{Emilie Frost}
\authornotemark[1]
\email{emilie.frost@offis.de}
\author{Malin Radtke}
\authornotemark[1]
\email{malin.radtke@offis.de}
\author{Sebastian Lehnhoff}
\email{sebastian.lehnhoff@offis.de}
\affiliation{%
  \institution{OFFIS - Institute for Information Technology}
  \streetaddress{Escherweg 2}
  \city{Oldenburg}
  \country{Germany}
  \postcode{26121}
}

\renewcommand{\shortauthors}{Oest, Frost, Radtke et al.}
\begin{abstract} 
 The increasing integration of renewable energy resources requires so-called smart grid services for monitoring, control and automation tasks. 
Simulation environments are vital for evaluating and developing innovative solutions and algorithms.
Especially in smart energy systems, we face a variety of heterogeneous simulators representing, e.g., power grids, analysis or control components and markets. 
The co-simulation framework mosaik can be used to orchestrate the data exchange and time synchronization between individual simulators. So far, the underlying communication infrastructure has often been assumed to be optimal and therefore, the influence of e.g., communication delays has been neglected.
This paper presents the first results of the project cosima, which aims at connecting the communication simulator OMNeT++ to the co-simulation framework mosaik to analyze the resilience and robustness of smart grid services, e.g., multi-agent-based services with respect to adaptivity, scalability, extensibility and usability. This facilitates simulations with realistic communication technologies (such as 5G) and the analysis of dynamic communication characteristics by simulating multiple messages. We show the functionality and benefits of cosima in experiments with 50 agents.
\end{abstract}



\maketitle
\begin{datamaterial}
The data and code used in this paper are available at
\url{https://gitlab.com/mosaik/examples/cosima/-/tags/cosima_paper_12_22}.
\end{datamaterial}
\section{Introduction}
 Intelligent electricity grids, so-called smart grids, need to be flexible, available, reliable and economically attractive \cite{Niesse2014}.
The transition of existing electricity generation to renewable, decentralized generation increases the complexity of the overall system, as the monitoring and control mechanisms should be able to deal with many individual plants and also with (e.g. weather-related) feed-in fluctuations \cite{Niesse2014, vogtsurvey2018}. 
Intelligent control and monitoring mechanisms in energy systems require the integration of \acf{ICT} \cite{vogtsurvey2018}. 
Here, the properties and behavior of the \ac{ICT} system (e.g., the topology of the communication network, protocols, communication latency, bandwidth, information security and reliability issues) influence the connected power system \cite{palensky_cosimulation_2017}.
For instance, the lack of bandwidth may lead to an increase in the overall message delay and strongly delayed monitoring messages may lead to a decrease in the observability of the power system. Significantly delayed control messages may also cause a decreased power system state \cite{Hassan2021}.
On the other hand, the power grid also has an influence on the \ac{ICT} system 
\cite{palensky_cosimulation_2017}.
Firstly, if there is a need for better observability of the power system state, sensors like \acp{PMU} typically increase their sampling rate and hence, create more traffic on the \ac{ICT} system \cite{Miller2017, Das2021}. Secondly, devices in the \ac{ICT} infrastructure (such as routers, switches or radio towers for mobile communication) are often directly connected to the power grid. Therefore, if a power outage occurs, the availability and capability of the communication network are restricted \cite{Volkova2019}.

Apart from monitoring and control tasks, we assume that more and novel smart grid services will be (partially) based on multi-agent systems (MAS) to address autonomy, self-organization and self-optimization for a more resilient and robust power system \cite{Loeser2022}.
Such \ac{MAS}-based services may perform various tasks in the power system (e.g., voltage control \cite{Wol19b}, decentral black start \cite{Stark21}, unit commitment \cite{Nie2015, Hinrichs17}) as logically or physically distributed algorithms.
\acp{MAS} can include numerous agents and depend on a period of extensive message exchange between individual agents for e.g., cooperative problem-solving. This can also involve the data exchange of complex objects (e.g., multiple schedules) between agents.
Therefore, they depend on the underlying \ac{ICT} infrastructure as the behavior of the MAS may be influenced by communication properties such as latency.
This assumption is made without loss of generality. Non-\ac{MAS} services in smart grids that are dependent on \ac{ICT} can also be significantly affected by communication deficits.
Because of the criticality of \acp{CPES}, systematic testing of new technologies is essential to ensure functionality, stability and safety. Due to the complexity and the interdisciplinary of the systems, the investigation of the behavior of \ac{CPES} is mostly done through co-simulations \cite{steinbrink_cpes_2019}. 

The co-simulation framework mosaik \cite{mosaik} allows the combination of different, heterogeneous simulators from different domains and thus also the creation of complex power grid scenarios with a large number of loads and generators. The framework mosaik has already been used in multiple smart grid-related projects. Therefore, a collection of smart grid components and tools have been developed and many of them are provided as free simulators in the mosaik ecosystem \cite{mosaik-ecosystem} or in the \ac{MIDAS} \cite{midas}, which also includes a semi-automatic scenario configuration tool. This reduces the time to develop new or extend existing smart grid scenarios in mosaik.
The framework mosaik facilitates the data exchange and the temporal synchronization of simulators. As the data between simulators are directly exchanged, mosaik-based simulations assume perfect communication conditions. 
In order to represent a more realistic communication behavior, external communication simulators, such as the framework OMNeT++ \cite{Varga2010, omnetpp}, need to be integrated into a mosaik-based simulation. 
Whereas most power grid simulators operate in time discrete simulations within mosaik (i.e., advancing simulation time in fixed discrete time intervals), communication simulators run in discrete event simulations (i.e., advancing in flexibly sized time intervals). 
The coupling of multiple simulators with heterogeneous scheduling algorithms and independent simulation clocks is therefore quite challenging \cite{Dede2015}. 
Previous coupling approaches based on mosaik 2 were limited to time discrete simulations. In order to capture the effects of a communication simulation, the mosaik simulation must propagate the simulation time in rather small time steps (e.g., seconds), which leads to more simulation time steps and to more calls from mosaik to OMNeT++ causing challenges to simulation efficiency \cite{Dede2015, veith_large-scale_2020}. 
The latest version of mosaik (3.0) supports event-based simulation \cite{ofenloch2022mosaikl}. This allows the coupling of energy system simulations with communication simulations since in this way different time scales of the systems
(typically in the range of several minutes for energy systems and milliseconds for communication systems) can be efficiently and performantly implemented \cite{palensky_applied_2017}. 

This paper presents the integration of the simulator OMNeT++  \cite{Varga2010} for (communication) network modeling into the co-simulation framework mosaik within the project \ac{cosima}\footnote{\url{https://cosima.offis.de/}},
which aims to analyze the interactions between power and communication systems and the resilience and robustness of smart grid services via co-simulation. 
Therefore, \ac{cosima} enables the simulation of messages in realistic communication infrastructures and technologies (e.g., \ac{LTE} or 5G), capture the dynamic effects of multiple, simultaneously transmitted messages and perform alterations on the \ac{ICT} topology during the simulation.  
Furthermore, \ac{cosima} enables to simulate the exchange of adaptable messages of numerous agents using mosaik~3 functionalities to conduct a scalable simulation. 
In addition, \ac{cosima} provides extensibility for further functionalities while maintaining high usability.

The paper is organized as follows:
In \autoref{sec:related_work} the related work is summarized. 
In \autoref{sec:description_cosima} we describe the architecture and functionality of \ac{cosima}. 
For this purpose, the components and the process of simulating a message dispatch are explained. 
In addition, the synchronization between mosaik and OMNeT++, scenarios and features of \ac{cosima} are explained. 
In \autoref{sec:evaluation} the functionality and scalability of the simulation are evaluated with experiments for a simple use case.
\autoref{sec:conclusion_future_work} concludes the paper and shows future work. 
\section{Related Work}
\label{sec:related_work}
The consideration of communication aspects for services in smart grids has already been the focus of several projects and many can be found in e.g., the surveys of Mets et al. \cite{mets_combining_2014} and Tan et al. \cite{Tan2019}.
One subcategory of smart grid services comprises (complex) distributed algorithms, such as coordinated \aclp{MAS} that can be used for virtual power plant management \cite{holker_communication_nodate, Oest21}, battery swarm storage management \cite{frost2020robust}, black-out restoration with distributed energy resources \cite{Stark21}, 
which rely on intensive message exchange in uncertain environments and are therefore, prone to communication delay, jitter and packet losses. This type of implementation of smart grid services may also introduce noticeable traffic on the communication system. 
For the purpose of abstract delay analysis, the communication can be directly integrated into the overlay topology of a \acs{MAS}, where agents sleep according to predefined end-to-end delays \cite{Stark21, frost2020robust}. This delay can be specified through a realistic communication simulation in EXata with concrete communication technologies, such as \ac{LTE}-Advanced \cite{Oest21}. 
Hence, these works lack a detailed communication simulation considering the dynamic delay of the traffic introduced by the simulated messages. 
The dynamics of mutual influence of simultaneous messages exchanged by agents in a \ac{MAS} was observed in an integrated \ac{MAS} and communication simulation with ns-3 for bitrate resources, but lack detailed simulation for latency \cite{holker_communication_nodate}. 
The project Intertwined \cite{Dorsch2016} aims at a reduction of end-to-end delay of \ac{MAS} communication by integrating an adaptive packet prioritization for messages of agents in a \ac{MAS} in order to guarantee voltage stability. This simulation environment is based on \ac{HiL} devices, which is not suitable for large-scale scenarios. 

Similarly, in other scenarios a joint study on \ac{ICT} and power systems has been conducted for further \ac{ICT} analysis (e.g., such as cyber-security).
For this, the co-simulation framework mosaik was used to synchronize data between a power system simulation and a Docker-based communication emulation for false data injections \cite{van_der_Velde21} or \ac{ICT} traffic and behavioral anomaly detection \cite{veith2020analyzing}.

Another possibility for coupling multiple analysis tools is presented in the project LarGo \cite{veith_large-scale_2020},  which incorporated PowerFactory, OMNET++, simona and Software-in-the-Loop-Applications based on Docker in a simulation with mosaik responsible for data exchange. 
However, the aforementioned works are based on communication emulation or \aclp{HiL}-integration, which are as well not suitable for large-scale scenarios. Furthermore, the co-simulation framework mosaik has been used in these projects in a time discrete simulation. As pointed out by the authors of \cite{veith_large-scale_2020}, performance improvements are expected when mosaik is used in an event-discrete simulation with communication simulators.
In addition, a power system solver like OpenDSS can be directly coupled to the OMNeT++ \cite{BHOR2016274}.
For this, it is noted that the continuous scheduling of OpenDSS is complex to couple with OMNeT++ and leads to causality errors, which requires to rollback and rerun part of the simulation.
In addition, the co-simulation capability of OpenDSS is limited to power system solving scenarios, since OpenDSS focuses on the exchange of power values \cite{opendss-protocoll}.  

Furthermore, there were efforts in coupling other co-simulation frameworks with communication simulators. An \ac{HLA}-coupling with OMNeT++ is described by Galle et al.\cite{HLA-OMNeT} in which the OMNeT++ scheduler is replaced by the rtiScheduler to incorporate messages from \ac{HLA}. This coupling has been used in smart grid scenarios using JADE as a \ac{MAS}-framework for simulating simple agent behavior and Simulink-based simulators to study distribution system restoration \cite{albagli_smart_2016}. Similarly, \ac{HLA} has been used in smart grid scenarios with the network simulator OPNET \cite{shum_hla_2014, Bottura, Georg14} and ns-2 \cite{Georg12}. 
\ac{HLA}-based simulations may have some drawbacks. 
Simulators in \ac{HLA} are described as powerful and autonomous, but the interaction between \ac{HLA}-based simulators must be set up by the user and can be quite time-consuming \cite{steinbrink_smart_2018}. In addition, to our knowledge, there are only the few projects using (pure) \ac{HLA} for co-simulation mentioned in \cite{albagli_smart_2016, shum_hla_2014, Bottura, Georg14, Georg12,WANG2022104008, NAN2011149, MAGOUA2022102529} in smart grid simulations, for which the code of the coupling and simulators are hardly available (anymore). Therefore, the lack of reusability of existing projects and the complexity of the coupling creates a barrier for using \ac{HLA}. 

The \ac{FMI} is another standard, which is used for co-simulation in smart grid scenarios with integration of communication simulation. \ac{FMI} is used to connect simulators defined as \acp{FMU}. It has been used with OPNET \cite{Oudart17}, ns-3 and pandapower \cite{gougeon-co-simulation-nodate}, ns-3 and GridLAB-D \cite{tsampasis_novel_2016}, OMNeT++ and Modelica and EMTP-EV for electricity simulation \cite{demazeau_multi-agent_2015}, Matlab and OMNeT++ \cite{neema_model-based_2014}. The simulation with \ac{FMI} requires an simulation master for time management and data exchange between \acp{FMU}\cite{neema_model-based_2014, tsampasis_novel_2016}. For this, in the project \ac{VirGIL} \cite{Virgil2}, ptolemy is used. In other projects \cite{demazeau_multi-agent_2015, hammoudi_smart_2021} \ac{HLA} is used as a master algorithm benefiting from the existing (and available) HLA-OMNeT++ coupling. 
Furthermore, \ac{FMI} is not adapted to discrete-event modeling and reactive systems \cite{oudart_model_2020}.

To summarize, there are numerous efforts to study the inter-dependency between the \ac{ICT} and power system with a variety of tool combinations in either a direct ad-hoc coupling or a co-simulation. 
However, \ac{HiL} approaches are not suited for large-scale scenarios, HLA-based simulation environments are lacking usability and \ac{FMI}-based simulation environments lack a master algorithm and are not suitable for discrete event simulations.
Furthermore, none of the aforementioned works are addressing modifications in the communication topology during the communication simulation in an integrated environment. 

\section{Description of cosima}
\label{sec:description_cosima}
The predominant goal of \ac{cosima} is the integration of the communication simulation framework OMNeT++ into mosaik-based simulations. 
Therefore, we will start by describing the co-simulation framework mosaik and OMNeT++ to create a basic understanding of the tools and their coupling challenges. We will then describe the general architecture of \ac{cosima}, its key components, and the synchronization between mosaik and OMNeT++. Then, we will provide an overview of the features of \ac{cosima} and how a \ac{cosima} scenario can be configured. At the end of this section, we will quickly compare our approach to other synchronization approaches.
\subsection{Fundamentals of mosaik and OMNeT++}
The co-simulation framework mosaik allows the combination of several simulators by coordinating the data exchange and time synchronization, thereby focusing on providing high usability and flexibility.
Simulators contain models that may represent real-world objects or systems (e.g., \ac{PV}-plant) and are able to create multiple instances of models (so-called entities).
The coupling of simulators (i.e., data exchange) and entity parameterization is defined in a mosaik scenario file. 
In order to be coupled with mosaik, simulators have to implement the mosaik simulator {interface}, which defines the entity parameterization, the data provided for other simulators and the simulation behavior. The simulation behavior is realized by \textit{stepping} the simulator through simulation time by incorporating the output information of other coupled simulators and simulation time information to perform an internal state transition of the simulator. 
At the end of a step, output data are created and the mosaik internal scheduler, which is responsible for the temporal synchronization of simulators, is informed about the time when this simulator should be stepped next. 
The mosaik scheduler was initially developed for time discrete simulation, where the size of the time between one simulator step to the next step (so-called step-size) was fixed. In version 3.0 of mosaik, it is possible to conduct discrete event simulations, which allow adaptive stepping. 
The scheduler is informed about the time steps by either the simulator itself or by the output of other simulators 
(i.e., when the information should be delivered to the receiving simulator). 
The scheduler of mosaik provides partial scheduling information to a simulator with, e.g., the \textit{max\_advance}-value. This value is determined by other (directly or indirectly) scheduled events for this (discrete event) simulator and indicates how far this simulator can advance in the simulation time without expecting new inputs. This functionality can help to define the synchronization between multiple discrete event simulators and avoids rolling back the simulation.

OMNeT++ is a discrete event-based C++ simulation framework and library for building network simulators. 
OMNeT++ provides a component architecture for simulation models. Modules can be connected to other modules, connections can have attributes such as delay and data rate and connected modules may eventually form a network. 
The data exchange between modules is realized through \textit{messages}, which can be used to represent frames or packets. Similarly to mosaik~3.0, the simulation of OMNeT++ modules is triggered by events, which can either be self-scheduled events or messages (which is a type of event). The OMNeT++ scheduler is responsible for maintaining and executing the events in the \ac{FES}, which defines the order of event execution.
The INET framework \cite{INeTinbook} extends OMNeT++ with models for the internet stack and many other internet-related protocols (e.g., Ethernet, \ac{IP}, \ac{UDP} and \ac{TCP}) and modules (e.g., routers, switches and hosts). 
INET also serves as a basis for other frameworks, such as SimuLTE \cite{SimuLTEinProceedings} and Simu5G (mobile network) \cite{Simu5GArticle}, which allows detailed communication network scenarios and simulations. 
\subsection{General Architecture}
The following introduces the general architecture of \ac{cosima} including its components. After explaining the main components on the mosaik and OMNeT++ side, an example of the process of sending a message is discussed. 
\subsubsection{Components}
In the description of the components, the components in mosaik are presented first: the \texttt{Communica\-tion\-Simulator}, which represents the interface from mosaik to OMNeT++. Similarly, in OMNeT++ exists the \texttt{Mosaik\-Sched\-uler} for the interface to mosaik, which is described afterwards. Therefore, the \ac{TCP} connection between mosaik and OMNeT++ is created via the \texttt{Communica\-tion\-Simulator} and the \texttt{Mosaik\-Sched\-uler}. 
This connection is used to transfer messages. 
Subsequently, further simulators and the associated instances in OMNeT++ are explained.

The \texttt{Communica\-tion\-Simulator} is a mosaik simulator and the interface between mosaik and OMNeT++ for other mosaik simulators. 
It creates and holds the \ac{TCP} connection to OMNeT++, receives messages 
{via mosaik from other mosaik simulators} and forwards those to OMNeT++. If messages from OMNeT++ are finished being simulated, the \texttt{Communication\-Simulator} receives and forwards those to the receiver simulator of the message in mosaik. 

The \texttt{Mosaik\-Sched\-uler} is implemented in OMNeT++ and is the interface to mosaik. It is inherited from the \texttt{cScheduler} in OMNeT++ and thus implements the standard functions of the scheduler in OMNeT++.  
Additionally, it holds the \ac{TCP} connection to mosaik. Whenever a message should be simulated in OMNeT++, the \texttt{Mosaik\-Sched\-uler} receives it from mosaik and inserts it into the \ac{FES}. This message is later executed as an event by the module corresponding to the sender. When a message has been simulated, the \texttt{Mosaik\-Sched\-uler} sends a message back to mosaik. 

With \ac{cosima}, the \texttt{ScenarioManager} module is provided in OMNeT++, with which it is possible to change the communication infrastructure in the modelled network in OMNeT++ from the mosaik side at runtime of the simulation, e.g. to disconnect or reconnect end devices. This functionality is demonstrated in \autoref{sec:evaluation}. 

There are also components that are simulators in mosaik and modules in OMNeT++. 
Since mosaik allows the combination of many heterogeneous simulators, diverse simulators can also be connected in the scenario in \ac{cosima}. Each simulator that sends messages to others is represented in OMNeT++ as a corresponding module. 
In the following example, these are the \texttt{Agent\-Sim\-ula\-tors}, because the messages exchanged between agents are simulated within OMNeT++. 
For each \texttt{Agent\-Sim\-ula\-tor}, an end device with an \texttt{AgentApp} is modeled in OMNeT++. For this, the end devices incorporate the internet protocols modeled in the INET framework according to the OSI model, which enables the \texttt{AgentApp} on the application layer to communicate according to the OSI model.
When a message from mosaik arrives in OMNeT++ and is inserted into the \ac{FES} by the \texttt{Mosaik\-Sched\-uler}, the event is executed for the \texttt{AgentApp} belonging to the sender of the message. This app is then responsible for sending the message to the module representing the receiver of the message in OMNeT++ over the OMNeT++ network. 
If a module in OMNeT++ receives a message, it is passed back to mosaik via the \texttt{Mosaik\-Sched\-uler}. The exact behavior of the \texttt{AgentApp} is explained in the following in detail with an example. 
\begin{figure}[htb!]
 \centering
  \includegraphics[width=\linewidth]{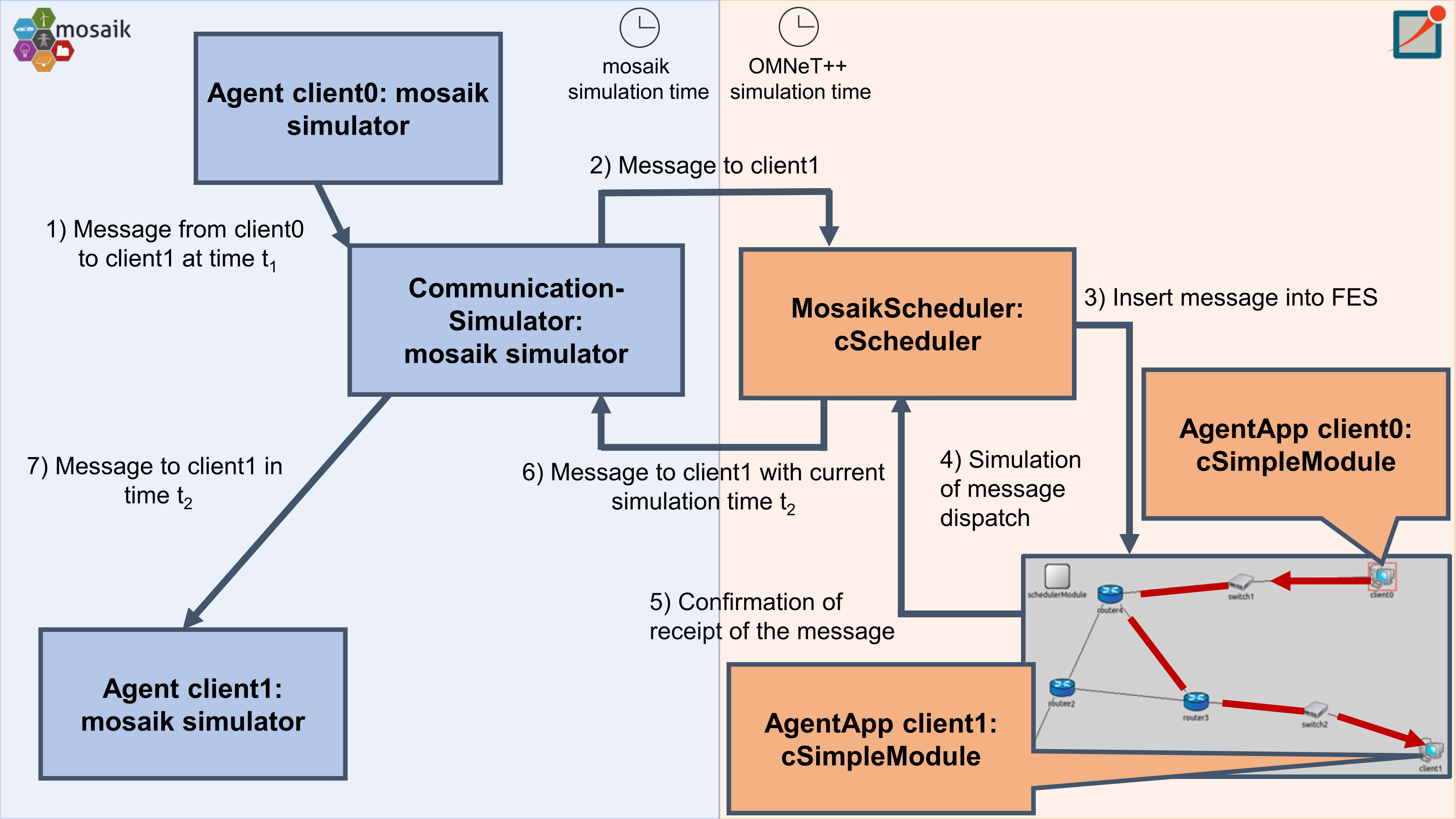}
  \caption{Example of a message exchange in \ac{cosima}}
  \label{fig:msg_exchange}
\end{figure}
\subsubsection{Interaction Example}\label{subsubsec:Process}
To illustrate the working principle and the interaction of the different components, an example of a message exchange between two agents is explained in the following. This example is illustrated in \autoref{fig:msg_exchange}. In this example, one \textit{client0} sends a message to \textit{client1}. The clients are both \texttt{Agent\-Sim\-ula\-tors} in mosaik. This message is sent via mosaik from \textit{client0} to the \texttt{Com\-mu\-ni\-cation\-Simu\-lator}, which then forwards it via the \ac{TCP}-Connection to the \texttt{Mosaik\-Sched\-uler}. The \texttt{Mosaik\-Sched\-uler} inserts it as an event. 
After a message from mosaik has been inserted into the \ac{FES} by the \texttt{Mosaik\-Sched\-uler}, the event is executed by the module corresponding to the \texttt{Agent\-Sim\-ula\-tor}. The corresponding module for \texttt{Agent\-Sim\-ula\-tors} in OMNeT++ are \texttt{AgentApps}, in this case the implemented \texttt{AgentApp} for \textit{client0}. The app is then responsible for sending the message via the OMNeT++ network to the module that represents the receiving \texttt{Agent\-Sim\-ula\-tor}. Therefore, the \texttt{AgentApp} for \textit{client0} sends the message to the receiver, which is the \texttt{AgentApp} for \textit{client1}. 
When a module receives a message, it informs the \texttt{Mosaik\-Sched\-uler} about the message. 
The \texttt{Mosaik\-Sched\-uler} then transmits the message over the \ac{TCP}-connection back to mosaik, where the \texttt{Communica\-tion\-Simulator} finally sends it to the receiver simulator (\textit{client1}). 
This message also includes the current simulation time of OMNeT++. Since the message arrives according to the communication infrastructure modeled in OMNeT++, message delays could be modeled in the simulation.
\subsection{Synchronization}
\label{subsec:synchronization}
OMNeT++ and mosaik both have their individual simulation clocks. A simulation clock is a variable defining the (global) simulation time. Therefore, a synchronization regarding the simulation time of both frameworks is necessary. 
Since mosaik synchronizes several simulators during co-simulations, the time synchronization is as well 
coordinated and executed by mosaik. 
This also includes the start and end time of the simulation. 
For the synchronization of the OMNeT++ and mosaik, we use different types of messages. 

The \textbf{InitialMessage} is sent to OMNeT++ at the beginning of the simulation in order to inform OMNeT++ about important parameters in the simulation on the mosaik side (e.g., step size, simulation end).

An \textbf{InfoMessage} is sent to OMNeT++ and contains the content from the sending simulator (in mosaik) for the receiving simulator (in mosaik). This type of message also includes the \texttt{max\_advance}\--value from mosaik indicating the next already scheduled event that can influence the \texttt{Communication\-Simu\-lator}. This is then used in OMNeT++ to define a synchronization point in order to enable mosaik to integrate further InfoMessages (or InfrastructureMessages).

A \textbf{SynchronizationMessage} is sent between mosaik and OMNeT++ in order to synchronize the clocks. This message contains different types to distinguish between the causes of this message. SynchronizationMessages of the type \texttt{MAX\_ADVANCE} are sent from OMNet++ to mosaik if the \texttt{max\_advance} synchronization point is reached. SynchronizationMessages of the type \texttt{WAITING} are mainly sent from mosaik to OMNeT++ to allow OMNeT++ to proceed with the simulation (but may also be sent from OMNeT++ to mosaik to inform mosaik about the current waiting of OMNeT++). This is to ensure that OMNeT++ is ahead or equal in simulation time and simulation results of OMNeT++ are fed into the mosaik simulation.
SynchronizationMessages of the type \texttt{TRANSMISSION\_ERROR} are sent from OMNeT++ to mosaik to inform the \texttt{Communication\_Simula\-tor} about lost messages (due to e.g., disconnects in the OMNeT++ network).

A \textbf{InfrastructureMessage} is sent to OMNeT++ and is used for modifications in the communication infrastructure in OMNeT++ such as disconnects or reconnects of routers, etc. 

\begin{figure}[htb!]
  \includegraphics[width=0.5\textwidth]{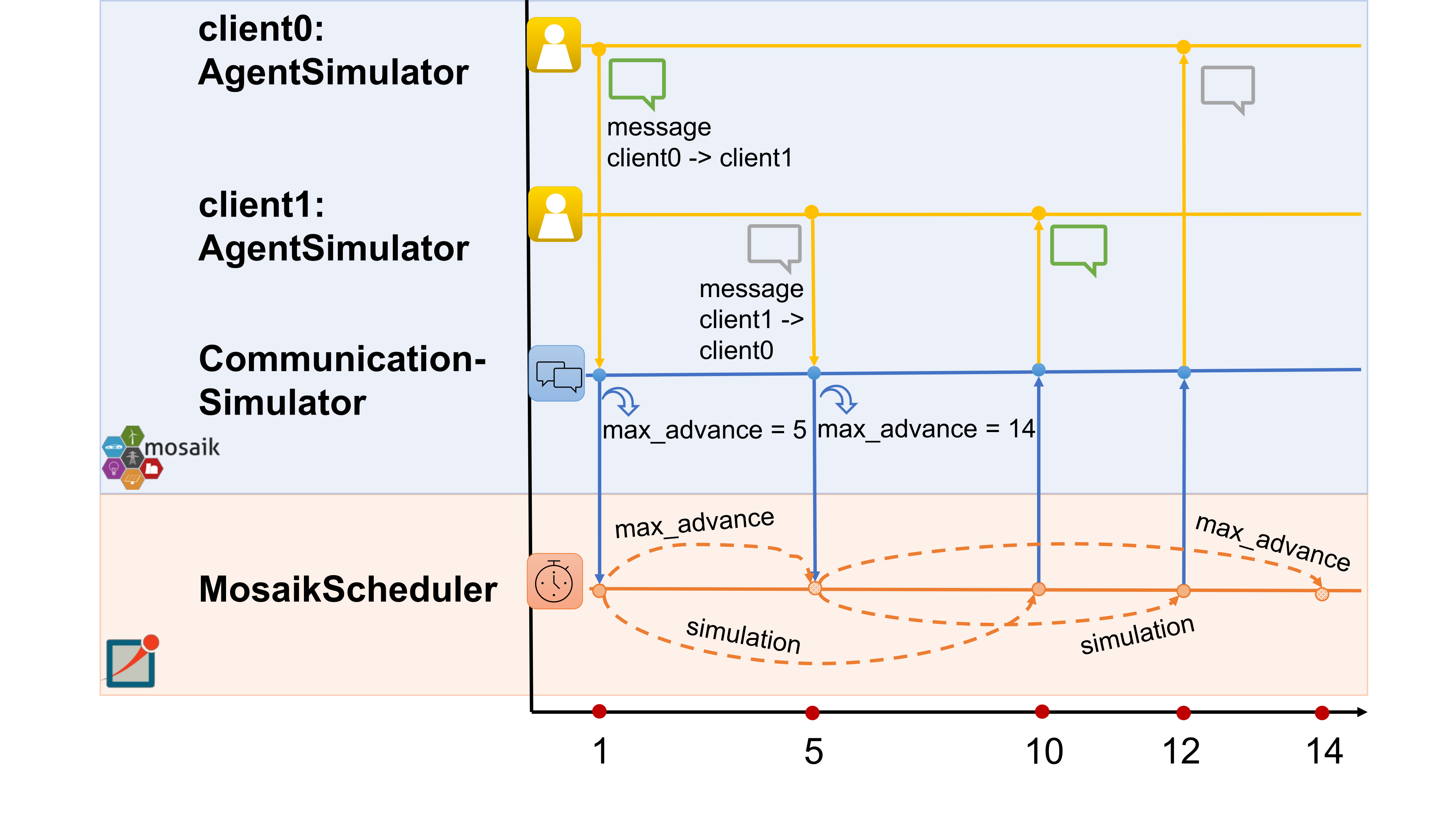}
  \caption{Synchronization example between OMNeT++ and mosaik}
  \label{fig:synchro}
\end{figure}

In the following, the synchronization process is explained in detail with an example. The process can be seen in \autoref{fig:synchro}.
In the beginning with $t = 1$, the process starts with an \texttt{Agent\-Sim\-ula\-tor} sending a message to another one. In this step, the first synchronization point is reached. In detail, \textit{client0} sends a message to \textit{client1}. To do this, it sends it to the \texttt{Communica\-tion\-Simulator} first, which then adds the current \texttt{max\_advance}-value from mosaik and forwards it to the \texttt{Mosaik\-Sched\-uler}. 
OMNeT++ takes this information of \texttt{max\_advance} as a synchronization point and connects to mosaik whenever the time of \texttt{max\_advance} was reached.
The value for \texttt{max\_advance} in this example is $t = 5$, which means that an event is scheduled in mosaik for $t = 5$. Within OMNeT++, the simulation of the message starts. As soon as the time reaches $t = 5$, a synchronization point in OMNeT++ is reached and the \texttt{Mosaik\-Sched\-uler} stops. 
Within mosaik, the event scheduled for $t = 5$ is then executed. In this case, this is a step from \textit{client1}, which sends a message to \textit{client0}. This message is sent to the \texttt{Communica\-tion\-Simulator}, which adds the new value for \texttt{max\_advance} (which is $t = 14$) and forwards it to the \texttt{Mosaik\-Sched\-uler}. 

The \texttt{Mosaik\-Sched\-uler} adds the event for \texttt{max\_advance} for $t = 14$ as new synchronization point and then continues to simulate. When the first message receives the module for \textit{client1}, another synchronization point is reached since the simulation of the message in OMNeT++ is completed and it can be sent back to mosaik. The \texttt{Mosaik\-Sched\-uler} then synchronizes with mosaik again and the \texttt{Communica\-tion\-Simulator} sends the message to its receiver (\textit{client1}). After this, the simulation in OMNeT++ continues and the \texttt{Mosaik\-Sched\-uler} synchronizes again with mosaik when the second message reaches its destination (here this is at $t = 12$) and at $t = 14$ (because of \texttt{max\_advance}).
To sum up, this example shows how the \texttt{max\_advance}\--value and the message insertion and extraction define synchronization points between mosaik and OMNeT++.
In order to synchronize via \texttt{max\_advance}, this value is transferred to OMNeT++ with messages from mosaik. The \texttt{Mosaik\-Scheduler} in OMNeT++ inserts an event into the \ac{FES} for the message it received from mosaik and furthermore adds an additional event for the time of \texttt{max\_advance}. Whenever an event indicating \texttt{max\_advance} occurs, OMNeT++ stops the simulation and synchronizes with mosaik, since an event is expected in mosaik at this time. 
If the time in OMNeT+ is more advanced than in mosaik, mosaik adapts to the simulation time of OMNeT++. This ensures that OMNeT++ does not advance too far. Also, no events from mosaik are skipped and therefore, we did not experience any casualty errors yet.

Whenever a message is sent from mosaik to OMNeT++, this message contains the current simulation time in mosaik. The \texttt{Com\-mu\-ni\-ca\-tion\-Simulator} checks whether the time in mosaik has not advanced further than the time in OMNeT++ when receiving a message from OMNeT++. In this way, the exchanged messages are used for the synchronization process.
\subsection{Scenarios}
Users of \ac{cosima} can create their own scenarios. The following features support this in an easy-to-use way. 
For this purpose, users can modify the configuration file provided with \ac{cosima}. 
In this configuration file, parameters can be defined like the number of agents, start and duration of the simulation, but also more complex functionalities. By simply setting the parameters in the configuration file, different scenarios can be easily executed. 
Therefore, the use of the configuration file increases usability.
The more complex functionalities of the scenario configuration are infrastructure changes, load or generation simulator integration and usage of predefined communication networks.
Clients, routers and switches can be disconnected and reconnected dynamically by integrating communication \textbf{infrastructure changes} into the mosaik scenario. Furthermore, \textbf{load or generation simulators} (as households, \ac{PV} plants) can be connected to agents. This can be used to consider the control of units by agents, which could receive power values and make adjustments or take control. The \textbf{predefined communication networks} can be used for a simulation. 

The networks contain different communication technologies and topologies. In detail, these are networks with different numbers of hosts in order to be able to map scenarios with different numbers of agents. The connections of the elements are, for example, wired Ethernet connections with different data rates of 10 or 100 Mbps or wireless connections as in \ac{LTE} networks. 
{These predefined scenario configurations support users in building their own scenarios.}
Additionally, it is easy to integrate own functionalities into the scenarios. This could e.g., be other agent or plant simulators or also self-developed networks for OMNeT++. {In this way, \ac{cosima} offers extensibility.}
This is supported by the adaptability of messages. The user can define its own messages to be sent via \ac{cosima}. These messages can also contain complex objects (e.g., combinations of lists and dictionaries).
\subsection{Features}
\label{subsec:features}
A lot of information about a simulation is stored in \ac{cosima}, as for instance the number of executed messages with the corresponding sender, receiver or message delay. It also contains information about \acs{CPU} and \acs{RAM} usage and the possibility to create illustrations based on the collected information. The stored information can also be easily supplemented with additional information that might be relevant for customized scenarios by the user. Thus, the user gets a comprehensive overview of the simulation run. 
With \ac{cosima} it is possible to run the simulation with \ac{IDE} or without to enable visualization or speed-up of simulation. 
Users can also adjust the \textbf{number of agents}, integrate \textbf{infrastructure changes} (dis- and reconnects of clients, routers and switches) and connect \ac{PV} plant or household simulators to agent(s). It is also possible to \textbf{add several OMNeT++ networks}.
\subsection{Comparison to other Synchronization Approaches}
In our approach, the \texttt{MosaikScheduler} replaces the default \texttt{cSche\-duler} in OMNeT++ in order to incorporate information from mosaik and enable the synchronization process. This is similar to the \ac{HLA}-OMNeT++ coupling \cite{HLA-OMNeT}. The usage of \texttt{max\_advance} to define synchronization points has been done in similar concepts for \ac{HLA} \cite{Georg14} and for \ac{FMI} \cite{tsampasis_novel_2016}. However, the introduction of changes in the communication infrastructure during the simulation and the associated challenges with lost messages adds new complexity to the synchronization.
For example, the \texttt{CommunicationSimulator} has to integrate the information about lost messages in order to proceed with the simulation.

\section{Evaluation}
\label{sec:evaluation}
As initially motivated, \ac{cosima} allows the co-simulation of use cases in modern energy systems with integrated communication simulation. 
In particular, we investigate the influence of the communication system properties, such as delay or communication disruptions on the behavior of individual agents and thus, the overall system. With this, we present the functionalities and benefits of \ac{cosima}.
\subsection{Description of the use case}
One possible use case of \ac{MAS} in distributed power systems is the provisioning of active power in a virtual power plant, which can be used to provide ancillary services.
In that use case, the goal is to determine a schedule for each participating \ac{DER} so that the resulting aggregated schedule of all \acp{DER} produces a global target power profile. 

This problem constitutes a distributed combinatorial optimization problem.
A fully distributed solution finding, without central components or hierarchies, can be achieved with the \ac{COHDA} \cite{hinrichs2014}. 

\ac{COHDA} is used for provisioning of active power \cite{hinrichs2013cohda} or the scheduling of energy units in virtual power plants \cite{niesse2016local, Oest21}. Furthermore, it is used in a wide variety of other use cases, such as demand side management \cite{holker_communication_nodate} or power system restoration \cite{Stark21}.
A set of \ac{DER} (represented as agents) aims to negotiate a solution to provide the global target power profile through message exchange within the neighborhood of the agents over a communication network. 

The neighborhoods of the agents are defined via a so-called overlay topology. which is a logical network built upon the underlying communication infrastructure. 
In \ac{COHDA}, a small-world topology according to Watts and Strogatz \cite{strogatz2001}, as shown in the appendix in \autoref{fig:overlay}, is typically used for best performance \cite{holker_communication_nodate}. 

The heuristic is executed via the three phases perceive, decide and act. 

In the \textbf{perceive phase}, the agent receives messages from its neighbors, updates its system state and waits for a defined waiting period. Then, the own solution is updated with the information from the solution of other agents in the received messages. If no new information arises, none of the next phases is performed.
However, if new information arises, the \textbf{decision phase} follows. Here the agent optimizes its own solution. 
Finally, the \textbf{act phase} follows. Here, the agent shares its system state by sending messages to the neighbors reachable through the overlay network. 

In the application of such distributed heuristics in power systems, communication between agents takes place via communication technologies, such as \ac{LTE} \cite{ghorbanian2019}. 
\subsection{Description of the simplified scenario}
The integration of communication simulation into the co-simulation with mosaik, enabled by \ac{cosima}, allows the investigation of the effects of communication parameters on the agent behavior in such \ac{MAS}. 
To illustrate this as an example, a simplified energy system scenario, based on the aforementioned scenario of the provisioning of active power using COHDA, was implemented. 

In order to show the scalability of \ac{cosima}, 50 agents were modeled as simulators in mosaik. 

Each of the agents is connected to a \ac{PV} system simulator and a household simulator in mosaik internally. These have been obtained from the mosaik ecosystem \cite{mosaik-ecosystem} and can be easily integrated into the scenario accordingly. Through the connections to these simulators, the agent receives the power values of its \ac{PV} plant and household. The agent is thus able to monitor (through incoming connections) and control (through outgoing connections) the generator or load. The agent is capable of aggregating the individual values in order to provide the resulting power product to its neighborhood. 

The goal of every agent is to form an association with a subset of the other agents in the network to provide a particular active power product above a given threshold (700\,W in this example). 

One of the agents starts exchanging messages with its neighbors. Inspired by \ac{COHDA}, the behavior of the agents (in the \texttt{step}-method of mosaik) is defined as follows: 
The agent first processes the received messages. These can be either power values from the connected (\ac{PV} and household) simulators or messages from its neighbors (agent simulators) reachable via the overlay network as shown in \autoref{fig:overlay}. 
The power values are stored in the agents' knowledge.
Incoming messages are first stored internally and then processed at the end of the waiting period. We chose a waiting period of 50\,ms so that messages can be received within a waiting period even when considering delays due to the communication network. 

At the end of the waiting period, the agent aggregates the power values it has knowledge about in its neighborhood. 

If the resulting value is above the defined threshold, a solution is found and the simulation terminates.
If not, the agent sends its updated knowledge to its neighbors. 

If the agent has already found a solution, meaning that it can provide an aggregate product together with its neighbors that is above the given threshold, nothing is done. \ac{COHDA} is here applied in a modified form, since in this, simplified implementation, the negotiation is terminated as soon as a solution has been found locally by an agent. If more than one agent has found a solution, any one of these can be selected. 

To simulate a scenario close to reality, a hybrid Ethernet-\ac{LTE}-based communication network was modeled in OMNeT++ as shown in \autoref{fig:omnetpp_network_lte}.
The agents (implemented as simulators in mosaik) are represented as end devices in OMNeT++. 
The modeled communication network consists of a core network containing a PDN gateway, a router and two base stations (eNodeBs). For the wired connections in the core network, we used the 10\,Gigabit/sec Ethernet link model \texttt{Eth10G} provided by INET. The modeled network covers about 360\,m in width and about 210\,m in height. 

The transmission delays result from the modeled cable length and distances between the end devices in the wireless network. 

As shown in \autoref{fig:omnetpp_network_lte}, the ScenarioManager provided by \ac{cosima} is also modeled in the network. This enables dynamic dis- and reconnection of communication components such as end devices and routers during simulation runtime from the mosaik side.
\subsection{Evaluation Results}
To investigate the influence of the communication infrastructure on the message exchange of the agents and thus on their behavior, the same scenario was executed twice -- once with and once without communication simulation. Then, a third experiment (also with communication simulation) was examined, in which an agent is dynamically switched off during the simulation.
The agents store which other agents are part of the solution and with which values they would contribute. The system states at specific points in time influence the behavior and actions of the agents. Therefore, these serve as a basis for discussion of the evaluation results. The concrete results of the evaluation can be found in \autoref{app:appendix}.

\subsubsection{Results without Communication Simulation}
When running the scenario without the communication simulation, the first agents found a solution after 150 milliseconds. It is important to note that due to the declared waiting time, the agents only check the stored messages every 50 ms. Thus, the solution, in this case, was found after three intervals. At this time, five agents already found a solution that exceeds the threshold. The system states of each agent with a solution after a solution was found are displayed in \autoref{fig:stateswithoutcomm}.
\begin{figure}[h]
\centering
\resizebox{0.4\textwidth}{!}{%
\begin{tikzpicture}
\begin{axis}[
    ybar stacked, ymin = 0,
    bar width=15pt,
    ymax = 1050,
    legend style={at={(1.2,1)},
     anchor=north,legend columns=1},
    ylabel={Power Values kW},
    xlabel={System States of Agents},
    symbolic x coords={client0, client2, client8, client18, client48},
    xtick=data,
    nodes near coords align={anchor=north},
    every node near coord/.style={
    },
    enlarge x limits=0.30,
    extra y ticks = 700,
    extra y tick style={grid=major,major grid style={ultra thick,draw=black,}, tick label style={font=\bfseries,
    anchor=north east, xshift=30pt, yshift=12pt}},
    extra y tick labels={$\begin{aligned}Thres-\\hold\end{aligned}$},
    ]
    \addplot[fill=firstcolor,draw=black!80] plot coordinates {(client0, 164) (client2, 164) (client8, 164) (client18, 164) (client48, 164)};
    \addplot[fill=secondcolor,draw=black!80] plot coordinates {(client0, 354) (client2, 354) (client8, 354) (client18, 0) (client48, 0)};
    \addplot[fill=thirdcolor,draw=black!80] plot coordinates {(client0, 0) (client2, 344) (client8, 0) (client18, 0) (client48, 0)};
    \addplot[fill=fourthcolor,draw=black!80] plot coordinates {(client0, 0) (client2, 0) (client8, 344) (client18, 0) (client48, 0)};
    \addplot[fill=fifthcolor,draw=black!80] plot coordinates {(client0, 354) (client2, 0) (client8, 0) (client18, 354) (client48, 354)};
    \addplot[fill=sixthcolor,draw=black!80] plot coordinates {(client0, 0) (client2, 0) (client8, 0) (client18, 344) (client48, 0)};
    \addplot[fill=seventhcolor,draw=black!80] plot coordinates {(client0, 0) (client2, 0) (client8, 0) (client18, 0) (client48, 344)};
     \addplot[fill=eightcolor,draw=black!80] plot coordinates {(client0, 164) (client2, 0) (client8, 0) (client18, 0) (client48, 164)};
  \legend{client0, client1, client2, client8, client17, client18, client48, client49}
  \end{axis}
\end{tikzpicture}
}
  \caption{System states of the agents without communication simulation}
  \label{fig:stateswithoutcomm}
\end{figure}
\subsubsection{Results with Communication Simulation without Disconnects}
When running the simulation with the communication simulation, the first solution can be found at the exact same time (150 milliseconds). However, the systems states of the agents are quite different and furthermore, only three instead of five agents found a solution at this time (only \textit{client0}, \textit{client2} and \textit{client8}). The system states after the first agent found its solution can be seen in \autoref{fig:stateswithcomm}. Compared with the results in \autoref{fig:stateswithoutcomm},
it is clear that due to the consideration of (simulated) delays several messages have not been received by some agents at this point.
Therefore, it is not possible for all five agents to find a solution at this time. Some of the agents did not receive the messages which would be sufficient to exceed the threshold. Also noticeable is that \textit{client18} and \textit{client48} did not even receive the initial request from \textit{client0}. Although they have received the power values of the systems, they do not even try to find a solution and, accordingly, the system state of these agents in the figure does not even contain their own share. 
\begin{figure}[h]
\centering
\resizebox{0.4\textwidth}{!}{%
\begin{tikzpicture}
\begin{axis}[
    ybar stacked, ymin = 0,
    bar width=15pt,
    ymax = 1050,
    legend style={at={(1.2,1)},
     anchor=north,legend columns=1},
    ylabel={Power Values kW},
    xlabel={System States of Agents},
    symbolic x coords={client0, client2, client8, client18, client48},
    xtick=data,
    nodes near coords align={anchor=north},
    every node near coord/.style={
    },
    enlarge x limits=0.30,
    extra y ticks = 700,
    extra y tick style={grid=major,major grid style={ultra thick,draw=black,}, tick label style={font=\bfseries,
    anchor=north east, xshift=30pt, yshift=12pt}},
    extra y tick labels={$\begin{aligned}Thres-\\hold\end{aligned}$},
    ]
    \addplot[fill=firstcolor,draw=black!80] plot coordinates {(client0, 164) (client2, 164) (client8, 164) (client18, 0) (client48, 0)};
    \addplot[fill=secondcolor,draw=black!80] plot coordinates {(client0, 354) (client2, 354) (client8, 354) (client18, 0) (client48, 0)};
    \addplot[fill=thirdcolor,draw=black!80] plot coordinates {(client0, 0) (client2, 344) (client8, 0) (client18, 0) (client48, 0)};
    \addplot[fill=fourthcolor,draw=black!80] plot coordinates {(client0, 0) (client2, 0) (client8, 344) (client18, 0) (client48, 0)};
    \addplot[fill=fifthcolor,draw=black!80] plot coordinates {(client0, 0) (client2, 0) (client8, 0) (client18, 0) (client48, 0)};
    \addplot[fill=sixthcolor,draw=black!80] plot coordinates {(client0, 0) (client2, 0) (client8, 0) (client18, 0) (client48, 0)};
    \addplot[fill=seventhcolor,draw=black!80] plot coordinates {(client0, 0) (client2, 0) (client8, 0) (client18, 0) (client48, 0)};
    \addplot[fill=eightcolor,draw=black!80] plot coordinates {(client0, 0) (client2, 0) (client8, 0) (client18, 0) (client48, 0)};
  \legend{client0, client1, client2, client8, client17, client18, client48, client49}
  \end{axis}
\end{tikzpicture}
}
  \caption{System states of the agents with communication simulation}
  \label{fig:stateswithcomm}
\end{figure}
\subsubsection{Results with Communication Simulation with Disconnects}
To investigate how the system behaves when agents fail due to communication problems and are thus not reachable for others anymore, the described functionality of \ac{cosima} was used to dis- and reconnect modules. In this case, \textit{client1} was disconnected after 10 milliseconds. This number was arbitrarily chosen between 1 and 50 in order to disconnect dynamically during the simulation but before \textit{client1} was able to send a message after its waiting period. 
This agent was used since it was part of the solutions found in the previous scenarios and thus appears to be relevant to the solution. 

In this scenario, it took the agents 250 milliseconds to find a solution, therefore 100 milliseconds longer than without the disconnect of \textit{client1}. The results can be seen in \autoref{fig:stateswithdisconnect}. The system state of \textit{client0} changed. This agent previously also had the values of \textit{client1} in its system state, which is now disconnected. Now values from the other two agents are stored here. The fact that the agent received the values of these agents only in this scenario and not in the ones before can be explained by the duration of the simulation since it took 100 milliseconds longer than before. Nevertheless, the values are not sufficient to reach the threshold. It is noticeable that this time, not \textit{client2} and \textit{client8} have found their solutions, but \textit{client18} and \textit{client48}. In the previous results, \textit{client1} was part of the solution of \textit{client2} and \textit{client8} and is therefore crucial for these agents to find the solution. 
Since they are in the neighborhood of \textit{client1}, which is not reachable anymore, these agents do not manage to find the solution. In the scenario without communication simulation, it can be seen that \textit{client1} is not included in the solutions of \textit{client18} and \textit{client48}. Thus, the disconnect does not directly affect these two agents. Since the simulation with the disconnect of \textit{client1} ran longer, because no other agents found a solution before, the time was sufficient for these agents to find enough messages to exceed the threshold.
\begin{figure}[h]
\centering
\resizebox{0.4\textwidth}{!}{%
\begin{tikzpicture}
\begin{axis}[
    ybar stacked, ymin = 0,
    bar width=15pt,
    ymax = 1050,
    legend style={at={(1.2,1)},
     anchor=north,legend columns=1},
    ylabel={Power Values kW},
    xlabel={System States of Agents},
    symbolic x coords={client0, client2, client8, client18, client48},
    xtick=data,
    nodes near coords align={anchor=north},
    every node near coord/.style={
    },
    enlarge x limits=0.30,
    extra y ticks = 700,
    extra y tick style={grid=major,major grid style={ultra thick,draw=black,}, tick label style={font=\bfseries,
    anchor=north east, xshift=30pt, yshift=12pt}},
    extra y tick labels={$\begin{aligned}Thres-\\hold\end{aligned}$},
    ]
    \addplot[fill=firstcolor,draw=black!80] plot coordinates {(client0, 164) (client2, 0) (client8, 0) (client18, 164) (client48, 164)};
    \addplot[fill=secondcolor,draw=black!80] plot coordinates {(client0, 0) (client2, 0) (client8, 0) (client18, 0) (client48, 0)};
    \addplot[fill=thirdcolor,draw=black!80] plot coordinates {(client0, 0) (client2, 0) (client8, 0) (client18, 0) (client48, 0)};
    \addplot[ fill=fourthcolor,draw=black!80] plot coordinates {(client0, 0) (client2, 0) (client8, 0) (client18, 0) (client48, 0)};
    \addplot[fill=fifthcolor,draw=black!80] plot coordinates {(client0, 354) (client2, 0) (client8, 0) (client18, 354) (client48, 354)};
    \addplot[fill=sixthcolor,draw=black!80] plot coordinates {(client0, 0) (client2, 0) (client8, 0) (client18, 344) (client48, 0)};
    \addplot[fill=seventhcolor,draw=black!80] plot coordinates {(client0, 0) (client2, 0) (client8, 0) (client18, 0) (client48, 344)};
    \addplot[fill=eightcolor,draw=black!80] plot coordinates {(client0, 164) (client2, 0) (client8, 0) (client18, 0) (client48, 164)};
  \legend{client0, client1, client2, client8, client17, client18, client48, client49}
  \end{axis}
\end{tikzpicture}
}
\caption{System states of the agents with communication simulation and disconnect of \textit{client1}}
\label{fig:stateswithdisconnect}
\end{figure}
\subsubsection{Discussion of the Evaluation Results}
The results show how the consideration of communication changes the system states of the agents. Due to the fact that messages arrive with delays, the system states are different and for some agents, it takes longer to find a solution. Since communication delays are present in real networks, it is necessary to study the agent systems under consideration of realistic communication settings, which is easily enabled by the usage of \ac{cosima}.

The results further show the extent to which the availability of agents affects the message exchanges. If relevant agents are no longer available, some solutions could not be found anymore and some are found later. Since it also occurs in reality that agents are no longer reachable (e.g., due to cyber attacks), it is also necessary to examine agent systems with regard to communication failures in order to ensure their robustness and resilience. For this purpose, \ac{cosima} can be used. 

In the scenario with the disconnect of \textit{client1}, a possible solution strategy could be to dynamically adapt the overlay topology so that the formerly neighboring agents of \textit{client1} are better connected after its failure. Dynamic topology adjustments could be a possible solution in this case as shown in \cite{holly2021dynamic}. In order to be able to recognize when a topology adaptation would be useful, appropriate consideration of communication simulation is necessary when studying agent systems. The suitability of \ac{cosima} for this purpose was demonstrated with the example.

The investigations with communication influences with \ac{cosima} differ from previous work, e.g. in \cite{Oest21, frost2020robust}, because \ac{cosima} does not consider static delays, since the delays depend on the situation in the communication infrastructure and thus change dynamically. It considered the fact that messages in the network can influence each other and the communication infrastructure is based on real technologies. Furthermore, the influences of the failure of the communication infrastructure can be examined under realistic conditions.

\section{Conclusion and Future Work}
\label{sec:conclusion_future_work}
Simulation environments are widely used to evaluate new solutions and algorithms for modern energy systems with integrated \ac{ICT} mechanisms. In this paper, we presented the 
coupling of the communication simulator OMNeT++ and the co-simulation framework mosaik in the project \ac{cosima}, 
which enables integrated communication network simulation in power grid scenarios in order to investigate interactions between both domains and especially analyze the resilience and robustness of \ac{ICT}-enabled grid applications. 
For this, the architecture, components, synchronization mechanisms, scenarios and features of \ac{cosima} were presented.  

Since \ac{cosima} enables the simple creation of own scenarios with the given features or the facilitated integration of own features, the extensibility is given. 
In the evaluation, we addressed the scalability by simulating scenarios containing 50 agents, which negotiate and aggregate power values to reach a given target power value. The simulation was conducted in three experiments. 
We have simulated experiments with OMNeT++, where we modeled a hybrid \ac{LTE}- and Ethernet-based network; both with and without the communication simulator. 
With this, we studied the influence of the communication simulation on the solution finding process and the state of the agents. 
Furthermore, we have also conducted an experiment in which we disconnected one agent during the simulation. This led to a different system state of other agents and to an increase in negotiation duration. 
With our experiments, we were able to show the functionality of \ac{cosima} with respect to non-functional requirements (i.e., usability and extensibility), and the influence of realistic communication simulation on \ac{MAS}-based services.

In future work, we plan to extend \ac{cosima} by integrating more options to manipulate the communication infrastructure from mosaik simulators during the simulation.
We further plan to include e.g., adapting the routing of packets, mimicking \ac{DoS} attacks by simulating traffic or integrating jammer in wireless scenarios. We also plan to develop an analysis module that integrates and connects information from both mosaik- and OMNeT++-based simulation results and statistics. 
Furthermore, we plan to make \ac{cosima} widely and easily accessible by extending the usability with, e.g., tutorials and further example scenarios.

Apart from extending the project with further functionalities, \ac{cosima} will serve as a basis for future experiments regarding robustness and resilience evaluations of (multiple) \ac{MAS}-based services. 
This may include evaluating new resilience strategies, such as overlay topology adaption and dynamic re-prioritization of individual services.
For this, we plan to extend scenarios to include further mosaik-based simulators in order to use e.g., pandapower to analyze the power system state, and other strategies regarding the handling of message losses, high delays and failure of agents or other end devices.

\appendix
\section{Appendices}
\label{app:appendix}
 
 \begin{figure}[H]
 \centering
  \includegraphics[width=\linewidth]{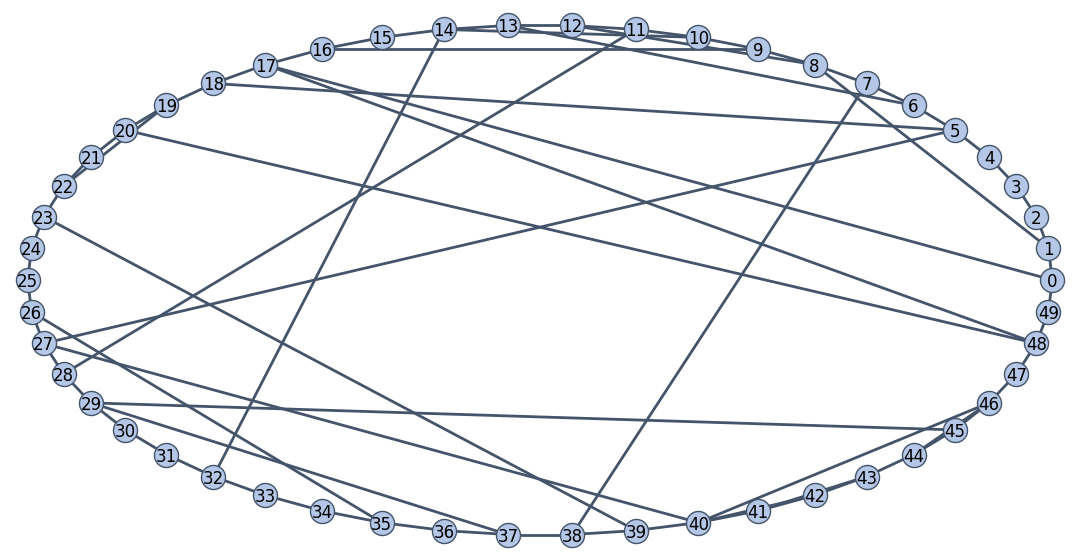}
  \caption{Overlay topology of the agents}
  \label{fig:overlay}
\end{figure}

\begin{figure}[h!]
    \centering
    \includegraphics[width=0.45\textwidth]{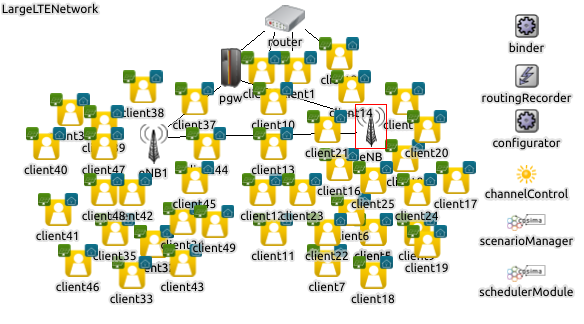}
    \caption{MultiCell LTE network in OMNeT++}
    \label{fig:omnetpp_network_lte}
\end{figure}

\begin{table}[h!]
\caption{System states of agents without communication simulation}
\label{tab:withoutcomm}
    \begin{tabular}{ccl}
    \toprule
    Agent & System State (Agent: Power in kW) \\
    \midrule
    client0 & client0: 164, client1: 354, client17: 354, client49: 164 \\
    client2 & client0: 164, client1: 354, client2: 344 \\
    client8 & client0: 164, client1: 354, client8: 344 \\
    client18 & client0: 164, client17: 354, client18: 344\\
    client48 & client0: 164, client17: 354, client48: 344, client49: 164\\
  \bottomrule
\end{tabular}
\end{table}

\begin{table}[h!]
\caption{System states of agents with communication simulation}
\label{tab:withcomm}
    \begin{tabular}{ccl}
    \toprule
    Agent & System State (Agent: Power in kW) \\
    \midrule
    client0 & client0: 164, client1: 354\\
    client2 & client0: 164, client1: 354, client2: 344 \\
    client8 & client0: 164, client1: 354, client8: 344 \\
    client18 & \\
    client48 & \\
  \bottomrule
\end{tabular}
\end{table}

\begin{table}[h!]
\caption{System states of agents with communication simulation and disconnect of \textit{client1}}
\label{tab:withcommdisconnect}
    \begin{tabular}{ccl}
    \toprule
    Agent & System State (Agent: Power in kW) \\
    \midrule
    client0 & client0: 164, client17: 354, client49: 164\\
    client2 & \\
    client8 & \\
    client18 & client0: 164, client17: 354, client18: 344 \\
    client48 & client0: 164, client17: 354, client48: 344, client49: 164\\
  \bottomrule
\end{tabular}
\end{table}


\section*{Abbreviations}
\begin{acronym}
\acro{COHDA}{Combinatorial Optimization Heuristic for Distributed Agents}
\acro{cosima}{\textbf{co}mmunication \textbf{sim}ulation with \textbf{a}gents}
\acro{CPES}[CPES]{Cyber-Physical Energy System}
\acrodefplural{CPES}[CPES]{Cyber-Physical Energy Systems}
\acro{CPU}{Central Processing Unit}
\acro{DER}{Distributed Energy Resource}
\acrodefplural{DER}[DERs]{Distributed Energy Resources}
\acro{DoS}{Denial of Service}
\acro{FES}{Future Event Set}
\acro{FMI}{Functional Mock-up Interface}
\acro{FMU}{Functional Mock-up Unit}
\acrodefplural{FMU}[FMUs]{Functional Mock-up Units}
\acro{HiL}{Hardware-in-the-Loop}
\acro{HLA}{High Level Architecture}
\acro{ICT}{In\-for\-ma\-tion and Com\-mu\-ni\-ca\-tion Tech\-nol\-o\-gy}
\acro{IDE}{Integrated Development Environment}
\acro{IP}{Internet Protocol}
\acro{LTE}{Long Term Evolution}
\acro{MAS}{Multi-Agent System}
\acrodefplural{MAS}[MAS]{Multi-Agent Systems}
\acro{MIDAS}{MultI-DomAin test Scenario}
\acro{PMU}{Phasor Measurement Unit}
\acrodefplural{PMU}[PMUs]{Phasor Measurement Units}
\acro{PV}{Photovoltaic}
\acro{RAM}{Random Access Memory}
\acro{TCP}{Transmission Control Protocol}
\acro{UDP}{User Datagram Protocol}
\acro{VirGIL}{Virtual Grid Integration Laboratory}

\end{acronym}
\bibliographystyle{ACM-Reference-Format}
\bibliography{main}

\end{document}